 \documentclass[letter]{aa} 
\usepackage{graphicx}
\usepackage{txfonts}
\usepackage{natbib}
\usepackage{longtable}
\usepackage{ulem}

\bibpunct[]{(}{)}{;}{a}{}{,}
\begin{document}
   \title{Detection of extragalactic argonium, ArH$^+$, toward PKS~1830$-$211}

   \author{Holger~S.~P. M{\"u}ller\inst{1}
           \and
           S{\'e}bastien Muller\inst{2}
           \and
           Peter Schilke\inst{1}
           \and
           Edwin A. Bergin\inst{3}
           \and
           John H. Black\inst{2}
           \and
           Maryvonne Gerin\inst{4}
           \and
           Dariusz C. Lis\inst{5,6}
           \and
           David A. Neufeld\inst{7}
           \and
           S{\"u}meyye Suri\inst{1}
          }

   \institute{I.~Physikalisches Institut, Universit{\"a}t zu K{\"o}ln,
              Z{\"u}lpicher Str. 77, 50937 K{\"o}ln, Germany\\
              \email{hspm@ph1.uni-koeln.de}
         \and
              Department of Earth and Space Sciences, Chalmers University of Technology, 
              Onsala Space Observatory, 43992 Onsala, Sweden 
         \and
              Department of Astronomy, The University of Michigan, 500 Church Street, 
              Ann Arbor, MI 48109-1042, USA
         \and
              LERMA, Observatoire de Paris, PSL Research University, CNRS, Sorbonne Universit{\'e}s, 
              UPMC Univ. Paris~06, {\'E}cole normale sup{\'e}rieure, 75005 Paris, France 
         \and
              LERMA, Observatoire de Paris, PSL Research University, CNRS, Sorbonne Universit{\'e}s, 
              UPMC Univ. Paris~06, 75014 Paris, France 
         \and
              Cahill Center for Astronomy and Astrophysics 301-17, 
              California Institute of Technology, Pasadena, CA~91125, USA 
         \and
              Department of Physics and Astronomy, Johns Hopkins University, Baltimore, MD 21218, USA
              }

   \date{Received 26 August 2015 / Accepted 22 September 2015}

  \abstract
{Argonium has recently been detected as a ubiquitous molecule in our Galaxy.  
Model calculations indicate that its abundance peaks at molecular fractions in the 
range of 10$^{-4}$ to 10$^{-3}$ and that the observed column densities require high 
values of the cosmic ray ionization rate. Therefore, this molecular cation may serve as 
an excellent tracer of the very diffuse interstellar medium (ISM), as well as an indicator 
of the cosmic ray ionization rate.}
{We attempted to detect ArH$^+$ in extragalactic sources to evaluate its diagnostic power 
as a tracer of the almost purely atomic ISM in distant galaxies.}
{We obtained ALMA observations of a foreground galaxy at $z = 0.89$ in the direction of the 
lensed blazar PKS~1830$-$211.}
{Two isotopologs of argonium, $^{36}$ArH$^+$ and $^{38}$ArH$^+$, were detected in 
absorption along two different lines of sight toward PKS~1830$-$211, known 
as the  SW and NE images of the background blazar. The argonium absorption is clearly 
enhanced on the more diffuse line of sight (NE) compared to other molecular species. 
The isotopic ratio $^{36}$Ar/$^{38}$Ar is $3.46 \pm 0.16$ toward the SW image, i.e., 
significantly lower than the solar value of 5.5.}
{Our results demonstrate the suitability of argonium as a tracer of the almost purely 
atomic, diffuse ISM in high-redshift sources. The evolution of the isotopic ratio with 
redshift may help to constrain nucleosynthetic scenarios in the early Universe.}
\keywords{quasars: absorption lines -- galaxies: ISM -- 
          astrochemistry -- galaxies: abundances -- 
          nucleosynthesis -- quasars: individual: \object{PKS 1830$-$211}
}

\authorrunning{H.~S.~P. M{\"u}ller et al.}

\maketitle
\hyphenation{con-tri-bu-tion}

%

\section{Introduction}
\label{intro}

Our knowledge of interstellar hydride species has been considerably increased by recent 
submillimeter missions, in particular the Heterodyne Instrument for the Far-Infrared 
(HIFI) on board \textit{Herschel} and the German REceiver At Terahertz frequencies 
(GREAT) on board the Stratospheric Observatory for Far-Infrared Astronomy (SOFIA), 
but also ground-based observatories, such as the Atacama Pathfinder EXperiment (APEX). 
The newly detected hydride species are SH$^+$ \citep{SH+_det_2011}, OH$^+$ 
\citep{OH+_det_2010}, H$_2$O$^+$ \citep{H2O+_det_2010}, H$_2$Cl$^+$ 
\citep{H2Cl+_det_2010}, HCl$^+$ \citep{HCl+_det_2012}, SH \citep{SH_det_2012}, and, 
most recently, ArH$^+$ \citep{ArH+_det_2013,ArH+_diff-ISM_det_2014}. 
OH$^+$ and H$_2$O$^+$ were also observed extensively in extragalactic sources 
\citep{det_OH+_Mrk231_2010,det_H2O+_M82_2010,OH_n^+_Arp220_2013}.

The large column densities of OH$^+$ with respect to H$_2$O$^+$ were explained by both 
molecules residing preferentially in the largely atomic, diffuse interstellar medium (ISM), 
with the largest fractional abundances at a molecular fraction of around 0.04, 
and their unexpectedly large column densities require cosmic ray ionization rates $\zeta$ 
considerably larger than in the dense ISM \citep{OH_n+_chemistry_2012,OH+_H2O+_H3+_2015}.

The ArH$^+$ cation was initially identified toward the Crab Nebula through its 
$J = 1 - 0$ and $2 - 1$ transitions, with OH$^+$ $1 - 0$ as the only other emission 
feature in the spectrum recorded with SPIRE (Spectral and Photometric Image REceiver) 
on board \textit{Herschel} \citep{ArH+_det_2013}. More recently, 
\citet{ArH+_diff-ISM_det_2014} reported on their early detections of the ArH$^+$ 
$J = 1 - 0$ transition in absorption toward six bright continuum sources using 
\textit{Herschel}-HIFI. The absorption patterns were unique in their appearances; 
ArH$^+$ showed up in all velocity components associated with diffuse foreground 
molecular clouds, but was conspicuously absent at velocities related to 
the sources themselves. The observations were  reproduced by models in which 
argonium is present only in the almost purely atomic diffuse ISM, with peak abundance 
at molecular fractions in the 10$^{-4}$ to 10$^{-3}$ range. Basically, high values of 
$\zeta$ favor its formation, consistent with analyses of OH$^+$ and 
H$_2$O$^+$, which trace slightly higher, but still low H$_2$ fractions 
(see, e.g., \citealt{OH_n+_chemistry_2012,OH+_H2O+_H3+_2015}), whereas molecular 
fractions $\gtrsim 10^{-3}$ favor its destruction by the reaction with 
H$_2$ to produce Ar and H$_3^+$. Although ArH$^+$ abhors molecular clouds, 
it does rely upon a small amount of H$_2$ for its existence. 
Moreover, with its specificity to the ISM with a molecular fraction of 
$\lesssim 10^{-3}$, it is a much better tracer of the almost purely atomic 
ISM than the 21~cm hyperfine structure line of atomic hydrogen, because 
the latter is also seen in the ISM with a molecular fraction of 0.01 or 
even larger than 0.5 \citep{ArH+_diff-ISM_det_2014}. Argonium may also 
be used to infer $\zeta$ in combination with other tracers.

Because argonium is ubiquitous in the Galactic diffuse ISM, the next step is 
to search for it in extragalactic sources. The $J = 1 - 0$ transition of 
$^{36}$ArH$^+$ at $617525.23 \pm 0.15$~MHz\footnote{From the CDMS catalog 
\citep{CDMS_1,CDMS_2}, see also section~\ref{obs}.} is close to a strong 
atmospheric water line ($5_{32} - 4_{41}$ near 620.7~GHz)\footnote{See, e.g., the 
JPL catalog \citep{JPL-catalog_1998}.\label{fn2}}, and the $^{38}$ArH$^+$ transition 
is less than 900~MHz lower in frequency. Therefore, ground-based observations 
of ArH$^+$ in absorption in nearby galaxies are very difficult. As \textit{Herschel} 
is no longer in operation and SOFIA has no receiver at these frequencies yet, 
searches toward galaxies at low to moderate redshifts are more promising, 
given the current instrumental capabilities.

The unnamed foreground galaxy at $z = 0.88582$ \citep{redshift_1996} in the 
direction of the blazar PKS~1830$-$211 is particularly well suited for such searches. 
The galaxy is lensing the background radiation coming from the blazar 
\citep{radio-flat-double_1988,lensing_1990}; two images, to the SW and NE 
of the blazar and separated by $1''$, are especially strong. Absorption spectra 
toward the SW and NE images probe two separate lines of sight across this 
foreground galaxy located on opposite sides of the nucleus. For most species, 
the absorption toward the SW image is much stronger than that toward the NE image, 
since it probes denser gas, with higher H$_2$ column densities. The blazar itself 
appears to be devoid of line-absorption or emission at radio frequencies. 
The line of sight toward the SW image of PKS 1830-211 is, to date, 
the extragalactic object with the largest number of molecules detected 
($\sim$40, see \citealt{survey_toward_PKS_2011,NH2_in_pks_etc_2014}).

The redshift $z = 0.89$ corresponds to a look-back time of $\sim$7.5~Gyr, 
when the Universe was slightly less than half its current age. At that time, 
the enrichment of the ISM with heavy elements was dominated by massive stars. 
This could lead to isotopic ratios of the elements different from those in the 
local galactic ISM. Molecular species may be much more promising in this 
regard than atoms, because of their more distinct spectroscopic patterns, 
e.g. different rotational patterns, caused by differences in the reduced masses. 
Indeed, isotopic ratios have been determined in the foreground galaxy for molecules 
containing moderately heavy elements such as Si, S, and Cl in addition to C, N, 
and O, and show some differences compared to their solar values 
\citep{isotopic-ratios_PKS_2006,survey_toward_PKS_2011,H2Cl+_extragal_2014}.

The Atacama Large Millimeter/submillimeter Array (ALMA) now offers high sensitivity 
and wide frequency coverage for spectroscopic studies. Already in the Early Science 
Cycle~0, despite the limited number of antennas and only four spectral tunings, 
a large number (17) of common interstellar species, such as CO, CH, H$_2$O, HCO$^+$, 
HCN, C$_2$H, and NH$_3$, were observed toward PKS~1830$-$211, including two new 
extragalactic detections, H$_2$Cl$^+$ \citep{H2Cl+_extragal_2014} and NH$_2$ 
(see \citealt{NH2_in_pks_etc_2014} for the presentation of the ALMA Cycle~0 
survey and overall results).

The present Letter reports the first extragalactic detection of $^{36}$ArH$^+$ and 
$^{38}$ArH$^+$ from ALMA observations toward PKS~1830$-$211.

\section{Observations and spectroscopic data}      
\label{obs}

\begin{figure}[ht!] \begin{center}
\includegraphics[width=8.8cm]{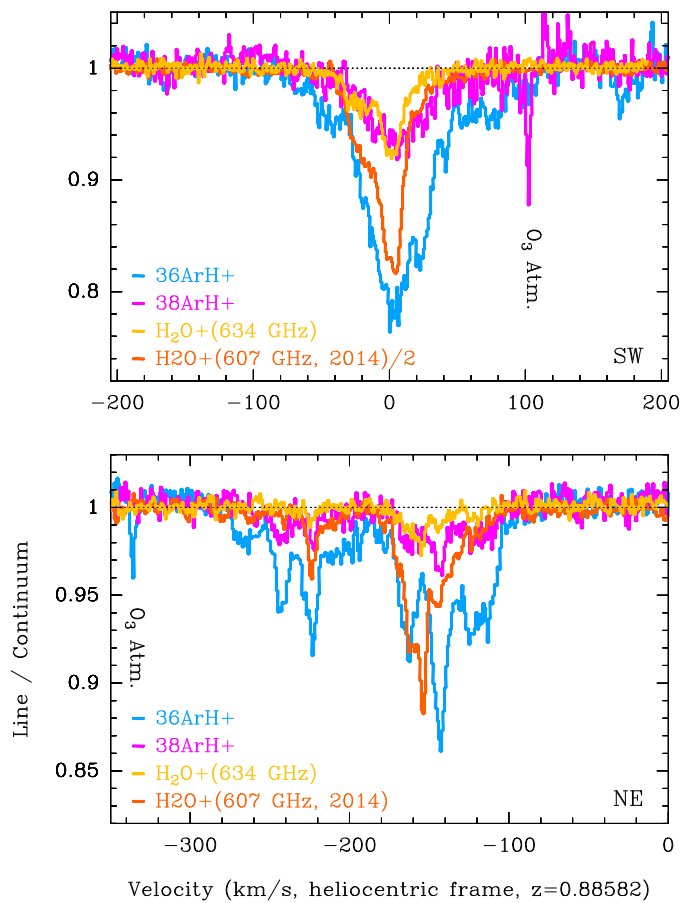}
\caption{Spectra of $^{36}$ArH$^+$ and $^{38}$ArH$^+$ toward PKS~1830$-$211 SW (top) 
         and NE (bottom) images, together with two \textit{para}-H$_2$O$^+$ fine 
         structure lines of the $1_{10} - 1_{01}$ transition; none of the transitions 
         shows hyperfine splitting.}
\label{fig:spec}
\end{center} \end{figure}


\begin{table*}
\begin{center}
\caption{Spectroscopic data$^a$ of the $J = 1 - 0$ transitions of ArH$^+$ isotopologs observed in the present study.}
\label{rest-freqs_etc}
\begin{tabular}[t]{llcrr@{}lccr@{}lr@{}l}
\hline \hline
Species  & \multicolumn{1}{c}{Rest frequency} &  $A_{ij}$$^b$        & \multicolumn{1}{c}{$E_{\rm low}$} 
  & \multicolumn{2}{c}{$\int \tau dv$(SW)$^c$} & $\Delta v$(SW)$^d$ & $v_{\rm off}$(SW)$^e$ &\multicolumn{2}{c}{$\int \tau dv$(SW)$^f$} &  \multicolumn{2}{c}{$\int \tau dv$(NE)$^g$} \\
         & \multicolumn{1}{c}{(MHz)}          & (10$^{-3}$~s$^{-1}$) & \multicolumn{1}{c}{(K)}           
  & \multicolumn{2}{c}{(km\,s$^{-1}$)}         & (km\,s$^{-1}$)     & (km\,s$^{-1}$)        & \multicolumn{2}{c}{(km\,s$^{-1}$)} & \multicolumn{2}{c}{(km\,s$^{-1}$)} \\
\hline
$^{36}$ArH$^+$ &  617525.23~(15) & 4.33 & 0.0 &  13&.65~(17)  & 57.3~(8) & 5.6~(4) &  15&.88~(21)   & 7&.61~(12) \\
$^{38}$ArH$^+$ &  616648.76~(8)  & 4.31 & 0.0 &   3&.90~(15)  & 57.3~(8) & 5.6~(4) &   4&.59~(21)   & 1&.68~(12) \\
$^{40}$ArH$^+$ &  615858.15~(5)  & 4.30 & 0.0 &    &          &          &         & $<$\,0&.42$^h$ &  &         \\
\hline
\end{tabular} 
\end{center}
\tablefoot{
$^a$ Numbers in parentheses are one standard deviation in units of the least significant figures. 
$^b$ Einstein $A$ value. 
$^c$ 
Velocity integrated optical depth toward the SW image from a single Gaussian fit. 
$^d$ Full width at half maximum of main SW image feature from that fit constrained to be equal for all three species. 
$^e$ Offset of main SW image centroid velocity with respect to the systemic velocity from that fit constrained to be equal for all three species. 
$^f$ Optical depth toward the SW image integrated over channels from $-$80 to +110~km\,s$^{-1}$. 
$^g$ Optical depth toward the NE image integrated over channels from $-$280 to $-$80~km\,s$^{-1}$. 
$^h$ Upper limit, 3$\sigma$ uncertainty. 
}
\end{table*}


Observations reported here were obtained with ALMA on 2015 May 19 (Cycle~2), 
in about 20~min on-source time under very good atmospheric conditions (precipitable 
amount of water vapor of $\sim$0.3~mm) and with receivers tuned to $\sim$327~GHz 
(Band~7)\footnote{At $z=0.89$, the frequencies of the $J=1-0$ line of ArH$^+$ 
isotopologs are redshifted close to the $5_{15} - 4_{22}$ atmospheric water line near 
325.2~GHz$^{\ref{fn2}}$.}. The three ArH$^+$ isotopologs were observed simultaneously 
in the same 1.875~GHz wide spectral window. Two additional spectral windows were centered 
at $\sim$337~GHz to cover the H$_2$O$^+$ $1_{10} - 1_{01}$, $J = 0.5 - 1.5$ line 
(rest frequency 634.27~GHz) and at $\sim$339~GHz (no lines were detected in this window, 
which was used to map the continuum). The bandpass response of the antennas 
was calibrated using the radio-bright quasar J1924$-$2914, and the gain solutions 
versus time were determined from observations of the quasar J1832$-$2039, within 1$^{\rm o}$ 
on the sky from PKS~1830$-$211. The overall standard calibration was applied by the ALMA 
pipeline.  One more step of self-calibration was performed and calibrated visibilities were 
then fit with the task UVMULTIFIT \citep{mar14} to extract the spectra toward the point-like 
SW and NE images of PKS~1830$-$211. The two images (separated by $\sim$1$\arcsec$) were well 
resolved by the synthesized beam of $\sim$0.5$\arcsec$. The velocity resolution is 
1.0 km\,s$^{-1}$ after Hanning smoothing. Two atmospheric lines due to O$_3$ (at 326.901 and 
327.845~GHz)$^{\ref{fn2}}$ show up in the spectra (Fig.~\ref{fig:spec}). 
They are easily identified and disentangled from absorption intrinsic to the source, since 
they occur at the same frequency in both SW and NE spectra, see Fig.~\ref{fig:overview}.

Spectroscopic data of ArH$^+$ transitions were taken from the CDMS \citep{CDMS_1,CDMS_2} 
and are summarized in Table~\ref{rest-freqs_etc}. An isotopic invariant fit of diverse 
rotational and rovibrational data was carried out, similar to \citet{ArD+_rot_1999}, 
to determine spectroscopic parameters of the molecule and to derive rest frequencies 
of rotational transitions of argonium isotopologs. The most relevant data for our 
observations are the $J = 1 - 0$ transitions of $^{36}$ArD$^+$, $^{38}$ArD$^+$, and 
$^{40}$ArD$^+$ from \citet{isos-ArD+_1-0_rot_1983}. As $^{40}$Ar is the most abundant 
Ar isotope on Earth, almost all of the remaining rotational 
\citep{ions_FIR_1987,ArH+_rot_high-J_1988,ArD+_rot_1999} and rovibrational data 
\citep{ArH+_IR_1982,NgH-D+_IR_1984,36-38ArH+_IR_1988} used in that calculation refer 
to $^{40}$ArH$^+$ or to $^{40}$ArD$^+$. The $^{40}$ArH$^+$ ground state dipole moment 
is 2.18~D \citep{hydride_cations_ai_2007}.

\section{Results and discussion}
\label{r_and_d}

\subsection{Column densities and abundances} 

Absorption signatures due to $^{36}$ArH$^+$ and $^{38}$ArH$^+$ were detected 
with good to very good signal-to-noise ratios toward both images of PKS~1830$-$211, 
SW and NE, see Fig.\ref{fig:spec}. The images and overview spectra are shown 
in Fig.~\ref{fig:overview} in the Appendix.

Toward the SW image, a simple fit with one Gaussian velocity component and 
the same width and velocity offset for both isotopologs yields a centroid 
velocity of $5.6 \pm 0.4$~km\,s$^{-1}$ and a full width at half maximum 
of $57.3 \pm 0.8$~km\,s$^{-1}$, which, to first order, reproduces well both 
$^{36}$ArH$^+$ and $^{38}$ArH$^+$ absorptions. Integrated opacities of 
$13.64 \pm 0.19$ and $3.89 \pm 0.16$~km\,s$^{-1}$ were derived for $^{36}$ArH$^+$ 
and $^{38}$ArH$^+$, respectively. We obtain only an upper limit of 
0.42~km\,s$^{-1}$ (at 3$\sigma$ confidence level) in the case of $^{40}$ArH$^+$. 
However, the ArH$^+$ line profile shows a slight deviation from a simple Gaussian, 
and fit results and values of integrated opacity over velocity channels from $-$80 
to +110~km\,s$^{-1}$ are given in a separate column in Table~\ref{rest-freqs_etc}. 
The weak velocity component at +170~km\,s$^{-1}$, previously only detected in the lines 
of HCO$^+$, HCN, and H$_2$O and all three strongly saturated near $v=0$~km\,s$^{-1}$ 
\citep{survey_toward_PKS_2011,NH2_in_pks_etc_2014}, is $-$ surprisingly $-$ also 
detected in the $^{36}$ArH$^+$ spectrum.

In contrast to the SW line of sight, where the absorption mostly resides in one 
bulky absorption feature (with the exception of the additional +170~km\,s$^{-1}$ 
component), the ArH$^+$ absorption profile toward the NE image shows a remarkable 
series of narrow (a few km\,s$^{-1}$) features spanning over $\sim$200~km\,s$^{-1}$, 
also seen, e.g., in the absorption profile of H$_2$O \citep{NH2_in_pks_etc_2014}.

With the assumptions that the rotational excitation temperature is equal to the radiation 
temperature of the cosmic microwave background at $z=0.89$, 5.14 K, and that the lines 
are optically thin, the total column densities of $^{36}$ArH$^+$ along the SW and NE lines 
of sight are $2.7 \times 10^{13}$~cm$^{-2}$ and $1.3 \times 10^{13}$~cm$^{-2}$, 
respectively, i.e., differing by only a factor $\sim$2. 
\cite{H2Cl+_extragal_2014} found a similar ratio of $\sim$3 for H$_2$Cl$^+$. 
In contrast, the H$_2$ column density, estimated from proxies such as CH, is about 
one order of magnitude higher toward SW \citep{NH2_in_pks_etc_2014}. 
The NE line of sight is known to be more diffuse and richer in atomic gas \citep{koo05}, 
thus the enhancement of ArH$^+$ on the NE line of sight relative to the SW one 
is not surprising. The detection of the ArH$^+$ absorption in the +170~km\,s$^{-1}$ 
velocity component toward the SW image suggests that this region has a low molecular 
fraction as well, or that time variations \citep{mul08} have made this feature stronger.

Since it is expected that ArH$^+$ traces a gas component with a very low H$_2$/H 
fraction, it would seem natural to compare its absorption profile with that of 
{\sc H\,i}. The {\sc H\,i} absorption has been previously observed toward the 
blazar\footnote{See also http://www.atnf.csiro.au/projects/askap/news\_commissioning 
\_10042014.jpg} \citep{HI_OH_PKS_1999,koo05}, but only at low angular resolution, 
not resolving the background continuum morphology. PKS~1830$-$211 is radically 
different at centimeter and submm wavelengths: the NE and SW images, which are 
point-like in the submm, are more extended in the cm; in addition, the emission 
from the pseudo Einstein ring can be seen in the cm regime \citep{lensing_1990}. 
Thus, comparison of spectra between such different frequencies is not 
straightforward.

Both H$_2$O$^+$ and ArH$^+$ trace the diffuse ISM component \citep{OH+_H2O+_H3+_2015}. 
Therefore, one may assume that both species show similar absorption features. This is not 
the case toward either of the PKS~1830$-$211 images, as can be seen in Fig.~\ref{fig:spec}; 
this also demonstrated in a correlation plot (Fig.~\ref{fig:scatter-plot}) in the Appendix. 
The 634~GHz line of H$_2$O$^+$ was observed simultaneously with ArH$^+$, while the 
607~GHz line, with a higher relative intensity and better signal-to-noise ratio, 
was observed in 2014 (Muller et al., in preparation). There appear to be no large 
variations in the H$_2$O$^+$ spectra, which were observed about one year apart. 
A poor correlation between the two molecular ions is also seen in Galactic sources 
\citep{ArH+_diff-ISM_det_2014}, suggesting that the diffuse ISM with a molecular fraction 
10$^{-4}$ to 10$^{-3}$ (as traced by ArH$^+$) is different from the diffuse ISM with a 
molecular fraction around 0.04 (as traced by H$_2$O$^+$). 
It is conceivable that at least part of ArH$^+$ absorption arises in the warm 
(8000~K) ionized medium (WIM), rather than entirely in the least molecular parts 
of the diffuse, neutral ISM. A detailed discussion of the chemistry of the WIM 
will be presented elsewhere.

\subsection{The $^{36}$Ar/$^{38}$Ar isotopic ratio at $z = 0.89$}  

The $^{36}$Ar/$^{38}$Ar ratio is $3.46 \pm 0.16$ (integrated over velocities 
from $-$80 to +110~km\,s$^{-1}$; $3.50 \pm 0.14$ from the Gaussian fit) 
toward the SW image, significantly lower than the solar value $5.50 \pm 0.01$ 
\citep{Ar-Kr_Xe_2011}, or the terrestrial value of $5.305 \pm 0.068$ 
\citep{composition_elements_2009}. Although $^{40}$Ar (from the radio-active decay 
of $^{40}$K) is dominant on Earth, we note that it plays only a minor role in the ISM. 
The $\alpha$ elements sulfur and silicon also show ratios about two times lower than 
their terrestrial values (see \citealt{isotopic-ratios_PKS_2006,survey_toward_PKS_2011}).
In contrast, the $^{35}$Cl/$^{37}$Cl ratio was found to be identical to the terrestrial 
value \citep{H2Cl+_extragal_2014}. The $^{36}$Ar/$^{38}$Ar ratio toward the NE image 
is $4.53 \pm 0.33$, between the SW ratio and the solar and terrestrial values and 
compatible with either within the uncertainties. The 3$\sigma$ limit for the 
$^{38}$ArH$^+$/$^{40}$ArH$^+$ ratio of > 11 is not constraining considering the solar 
ratio of $\sim$615 \citep{Ar-Kr_Xe_2011}, which is assumed to be close to the value 
in the local ISM.

\citet{evolution_iso-ratios_2006} calculated nucleosynthesis yields of core-collapse
supernovae (SNe) and hypernovae (HNe, which are more energetic by an order of
magnitude) to predict the evolution of isotopic ratios in the Milky Way, SNe and HNe 
being the dominant contributors to elements from Na to Fe with the possible exception 
of a few selected isotopes. They presented yields for different progenitor masses and
metallicities, and subsequently incorporated the results into models for the enrichment 
of the Milky Way \citep{evolution_iso-ratios_2011}.  While the nucleosynthetic yields 
for many elements, including $^{36}$Ar and $^{38}$Ar, show a non-monotonic dependence 
on progenitor mass, the predicted $^{36}$Ar/$^{38}$Ar ratio tends to decline with 
increasing metallicity and is typically smaller for HNe than for SNe. Nevertheless, 
even  for the highest metallicities considered in the Milky Way enrichment models 
($Z = 0.02$, or solar metallicity), the predicted $^{36}$Ar/$^{38}$Ar ratio is larger 
than the observed ratio in either the Sun or the PKS~1830$-$211 absorber. 
There are several uncertain assumptions in the nucleosynthesis and Galactic enrichment 
models, including the importance of convective mixing in the progenitor stars and 
the relative frequency of HNe; observations of isotopic ratios in diverse environments, 
such as  those presented here, promise to provide valuable constraints for future models.
There is no evidence for low metallicity in the foreground galaxy, but the exact value 
cannot be determined at present.

\section{Conclusion}
\label{conclusion}

We have detected for the first time extragalactic argonium in both of its 
important interstellar isotopologs. Therefore, ArH$^+$ is not only a good tracer 
of the almost purely atomic diffuse ISM in the Milky Way, but it is also suitable 
for investigations of extragalactic sources, possibly to a similar extent as OH$^+$ 
or H$_2$O$^+$. It will be interesting to search for ArH$^+$ in a selection of 
galaxies at moderately low to higher redshifts to obtain a more complete picture 
of the atomic component of the extragalactic ISM and to look for possible trends 
in the $^{36}$Ar/$^{38}$Ar ratio with redshift. 
The Si, S, Cl, and Ar isotopic ratios appear to be promising probes into 
the nucleosynthesis in the early Universe because their yield ratios depend 
strongly on the type of supernova, and all four elements form molecules 
that have been detected even in distant galaxies. 
It will be interesting to see if these ratios correlate with $\zeta$ as both 
aspects are affected by SN activity. 
Searches for argonium in nearby galaxies may be more promising with the 
potential future GREAT receiver on board SOFIA. Combined observations of 
ArH$^+$ with OH$^+$, H$_2$O$^+$, and HF or CH will provide insight 
into the partitioning of atomic and molecular hydrogen and into the 
cosmic ray ionization rate. Such an analysis is currently under way 
for PKS~1830$-$211 and will be published separately.


\begin{acknowledgements}
This paper makes use of the following ALMA data: ADS/JAO.ALMA$\#$2013.1.00296.S. 
ALMA is a partnership of ESO (representing its member states), NSF (USA) and 
NINS (Japan), together with NRC (Canada) and NSC and ASIAA (Taiwan) and KASI 
(Republic of Korea), in cooperation with the Republic of Chile. The Joint ALMA 
Observatory is operated by ESO, AUI/NRAO and NAOJ. The present investigations 
have been supported by the Deutsche Forschungsgemeinschaft (DFG) in the framework 
of the collaborative research grant SFB~956, projects A4 and C3. Support for 
this work was also provided by NASA through an award issued by JPL/Caltech. 
Our research benefited from NASA's Astrophysics Data System (ADS).
\end{acknowledgements}



\onecolumn

\begin{appendix}

\section{Complementary figures}


\begin{figure}[ht!] \begin{center}
\includegraphics[width=15cm]{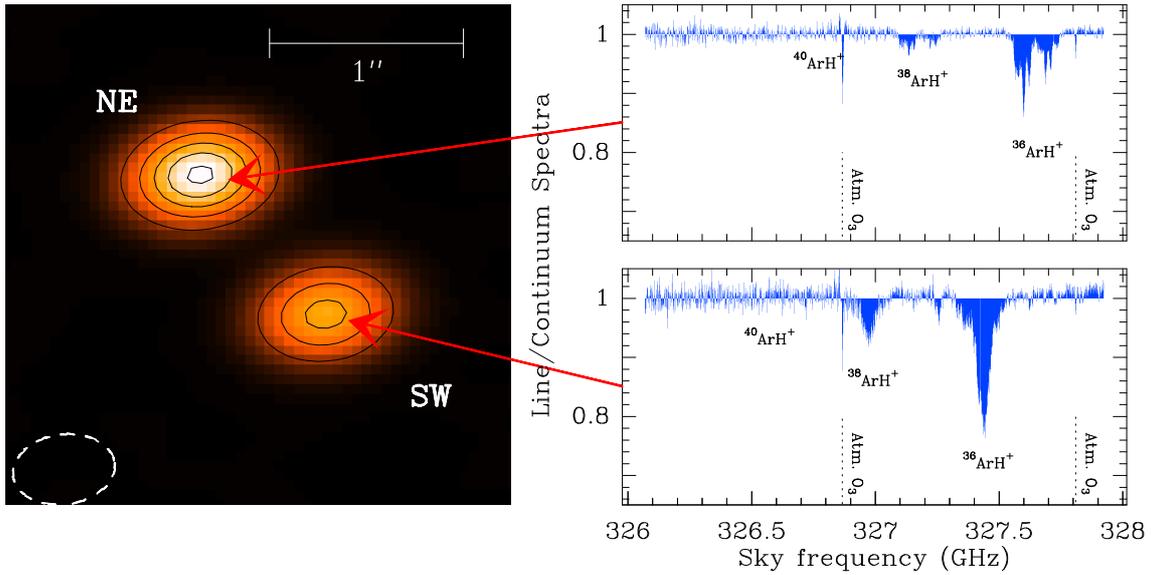}
\caption{Map of the 339~GHz continuum emission of PKS~1830$-$211 showing the two 
         resolved lensed images of the blazar (left). 
         Overview of the ArH$^+$ absorption spectrum toward PKS~1830$-$211 SW (bottom right) 
         and NE (top right) from the current observations. Expected positions of the main 
         absorption feature are indicated for $^{36}$ArH$^+$, $^{38}$ArH$^+$, and 
         $^{40}$ArH$^+$; the last is not detected. The noise increases toward lower 
         frequencies because of the proximity of the atmospheric water line near 325.2~GHz.}
\label{fig:overview}
\end{center} \end{figure}

\begin{figure}[hb!] \begin{center}
\includegraphics[width=10cm]{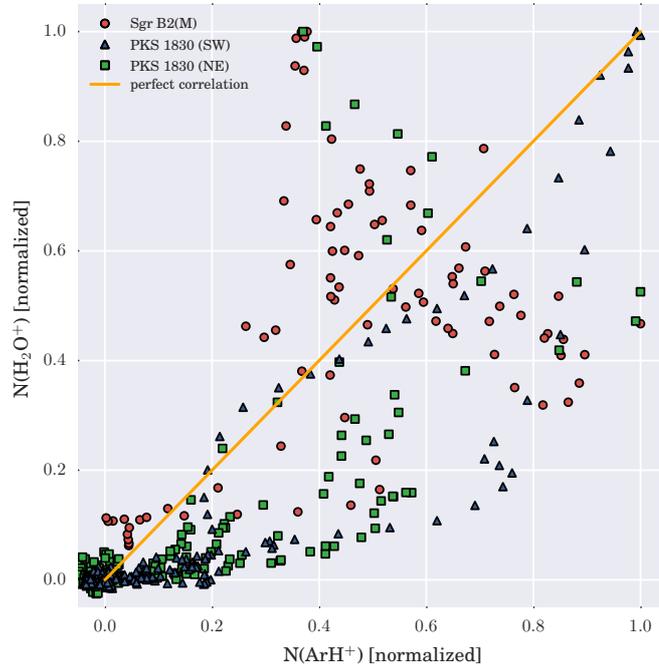}
\caption{Normalized correlation plot of H$_2$O$^+$ versus $^{36}$ArH$^+$ toward PKS~1830$-$211 
         SW, NE, and Sagittarius~B2(M). The column densities of the two species have been 
         normalized by the maximum values for each component and for each velocity channel. 
         A perfect correlation (i.e. just a constant scaling factor between the two species) 
         would result in clustering along the yellow line. The distributions of the two cations 
         are poorly correlated toward the sources. The PKS~1830$-$211 data are from this work 
         and from Muller et al., in preparation for $^{36}$ArH$^+$ and H$_2$O$^+$, respectively; 
         the corresponding Sagittarius~B2(M) data are from \citet{ArH+_diff-ISM_det_2014} and 
         \citet{H2O+_o-p_2013}.}
\label{fig:scatter-plot}
\end{center} \end{figure}

\end{appendix}


\begin{thebibliography}{}

\bibitem[Barlow et al.(2013)]{ArH+_det_2013} 
Barlow, M.~J., Swinyard, B.~M., Owen, P.~J., et al. 
2013, Science, 342, 1343 

\bibitem[Berglund \& Wieser(2011)]{composition_elements_2009}
Berglund, M; \& Wieser, M.~E.\
2011, Pure Appl. Chem., 83, 397

\bibitem[Bowman et al.(1983)]{isos-ArD+_1-0_rot_1983} 
Bowman, W.~C., Plummer, G.~M., Herbst, E., \& de Lucia, F.~C.\ 
1983, \jcp, 79, 2093 

\bibitem[Brault \& Davis(1982)]{ArH+_IR_1982} 
Brault, J.~W., \& Davis, S.~P.\ 
1982, \physscr, 25, 268 

\bibitem[Brown et al.(1988)]{ArH+_rot_high-J_1988} 
Brown, J.~M., Jennings, D.~A., Vanek, M., Zink, L.~R., \& Evenson, K.~M.  
1988, J. Mol. Spectrosc., 128, 587 

\bibitem[Cheng et al.(2007)]{hydride_cations_ai_2007} 
Cheng, M., Brown, J.~M., Rosmus, P., et al. 
2007, \pra, 75, 012502 

\bibitem[Chengalur et al.(1999)]{HI_OH_PKS_1999} 
Chengalur, J.~N., de Bruyn, A.~G., \& Narasimha, D. 
1999, \aap, 343, L79 

\bibitem[De Luca et al.(2012)]{HCl+_det_2012}
De Luca, M., Gupta, H., Neufeld, D., et al.,
2012, \apjl, 751, L37

\bibitem[Filgueira \& Blom(1988)]{36-38ArH+_IR_1988} 
Filgueira, R.~R., \& Blom, C.~E. 
1988, J. Mol. Spectrosc., 127, 279 

\bibitem[Gonz{\'a}lez-Alfonso et al.(2013)]{OH_n^+_Arp220_2013} 
Gonz{\'a}lez-Alfonso, E., Fischer, J., Bruderer, S., et al. 
2013, \aap, 550, A25 

\bibitem[Hollenbach et al.(2012)]{OH_n+_chemistry_2012} 
Hollenbach, D., Kaufman, M.~J., Neufeld, D., et al. 
2012, ApJ, 754, 105 

\bibitem[Indriolo et al.(2015)]{OH+_H2O+_H3+_2015} 
Indriolo, N., Neufeld, D.~A., Gerin, M., et al. 
2015, \apj, 800, 40 

\bibitem[Johns(1984)]{NgH-D+_IR_1984} 
Johns, J.~W.~C. 
1984, J. Mol. Spectrosc., 106, 124 

\bibitem[Kobayashi et al.(2006)]{evolution_iso-ratios_2006} 
Kobayashi, C., Umeda, H., Nomoto, K., Tominaga, N., \& Ohkubo, T. 
2006, \apj, 653, 1145 

\bibitem[Kobayashi et al.(2011)]{evolution_iso-ratios_2011} 
Kobayashi, C., Karakas, A.~I., \& Umeda, H. 
2011, \mnras, 414, 3231 

\bibitem[Koopmans \& de Bruyn(2005)]{koo05}
Koopmans, L. V. E. \& de Bruyn, A. G.
2005, \mnras, 360, L6

\bibitem[Lis et al.(2010)]{H2Cl+_det_2010} 
Lis, D.~C., Pearson, J.~C., Neufeld, D.~A., et al. 
2010, \aap, 521, L9 

\bibitem[Liu et al.(1987)]{ions_FIR_1987} 
Liu, D.-J., Ho, W.-C., \& Oka, T. 
1987, \jcp, 87, 2442 

\bibitem[Mart\'i-Vidal et al.(2014)]{mar14}
Mart\'i-Vidal, I., Vlemmings, W., Muller, S., \& Casey S. 
2014, \aap, 563, 136

\bibitem[Menten et al.(2011)]{SH+_det_2011} 
Menten, K.~M., Wyrowski, F., Belloche, A., et al. 
2011, \aap, 525, A77 

\bibitem[Muller et al.(2006)]{isotopic-ratios_PKS_2006} 
Muller, S., Gu{\'e}lin, M., Dumke, M., Lucas, R., \& Combes, F. 
2006, \aap, 458, 417 

\bibitem[Muller \& Gu\'elin(2008)]{mul08}
Muller, S. \& Gu\'elin, M.
2008, \aap, 491, 739

\bibitem[Muller et al.(2011)]{survey_toward_PKS_2011} 
Muller, S., Beelen, A., Gu{\'e}lin, M., et al. 
2011, \aap, 535, A103 

\bibitem[Muller et al.(2014a)]{H2Cl+_extragal_2014} 
Muller, S., Black, J.~H., Gu{\'e}lin, M., et al. 
2014a, \aap, 566, L6 

\bibitem[Muller et al.(2014b)]{NH2_in_pks_etc_2014} 
Muller, S., Combes, F., Gu{\'e}lin, M., et al. 
2014b, \aap, 566, A112 

\bibitem[M{\"u}ller et al.(2001)]{CDMS_1}
M{\"u}ller, H.~S.~P., Thorwirth, S., Roth, D.~A.,
\& Winnewisser, G.
2001, A\&A, 370, L49

\bibitem[M{\"u}ller et al.(2005)]{CDMS_2}
M{\"u}ller, H.~S.~P., Schl{\"o}der, F., Stutzki, J.,
\& Winnewisser, G.
2005, J. Mol. Struct, 742, 215

\bibitem[Neufeld et al.(2012)]{SH_det_2012} 
Neufeld, D.~A., Falgarone, E., Gerin, M., et al. 
2012, \aap, 542, L6 

\bibitem[Odashima et al.(1999)]{ArD+_rot_1999} 
Odashima, H., Kozato, A., Matsushima, F., Tsunekawa, S., \& Takagi, K. 
1999, J. Mol. Spectrosc., 195, 356 

\bibitem[Ossenkopf et al.(2010)]{H2O+_det_2010} 
Ossenkopf, V., M{\"u}ller, H.~S.~P., Lis, D.~C., et al. 
2010, \aap, 518, L111 

\bibitem[Pickett et al.(1998)]{JPL-catalog_1998} 
Pickett, H.~M., Poynter, R.~L., Cohen, E.~A., et al. 
1998, J. Quant. Spectrosc. Radiat. Transfer, 60, 883

\bibitem[Pramesh Rao \& Subrahmanyan(1988)]{radio-flat-double_1988} 
Pramesh Rao, A., \& Subrahmanyan, R. 
1988, \mnras, 231, 229 

\bibitem[Schilke et al.(2013)]{H2O+_o-p_2013} 
Schilke, P., Lis, D.~C., Bergin, E.~A., Higgins, R., \& Comito, C. 
2013, J. Phys. Chem. A, 117, 9766 

\bibitem[Schilke et al.(2014)]{ArH+_diff-ISM_det_2014} 
Schilke, P., Neufeld, D.~A., M{\"u}ller, H.~S.~P., et al. 
2014, \aap, 566, A29 

\bibitem[Subrahmanyan et al.(1990)]{lensing_1990} 
Subrahmanyan, R., Narasimha, D., Pramesh Rao, A., \& Swarup, G. 
1990, \mnras, 246, 263 

\bibitem[van der Werf et al.(2010)]{det_OH+_Mrk231_2010} 
van der Werf, P.~P., Isaak, K.~G., Meijerink, R., et al. 
2010, \aap, 518, L42 

\bibitem[Vogel et al.(2011)]{Ar-Kr_Xe_2011} 
Vogel, N., Heber, V.~S., Baur, H., Burnett, D.~S., \& Wieler, R. 
2011, \gca, 75, 3057 

\bibitem[Wei{\ss} et al.(2010)]{det_H2O+_M82_2010} 
Wei{\ss}, A., Requena-Torres, M.~A., G{\"u}sten, R., et al. 
2010, \aap, 521, L1 

\bibitem[Wiklind \& Combes(1996)]{redshift_1996} 
Wiklind, T., \& Combes, F. 
1996, \nat, 379, 139 

\bibitem[Wyrowski et al.(2010)]{OH+_det_2010} 
Wyrowski, F., Menten, K.~M., G{\"u}sten, R., \& Belloche, A. 
2010, \aap, 518, A26 


\end{thebibliography}
\end{document}